\documentclass[conference]{IEEEtran}
\usepackage[english]{babel}
\usepackage{cite}
\usepackage{color}
\usepackage{xcolor}

\usepackage{booktabs}
\usepackage{csquotes}
\usepackage{tablefootnote}
\usepackage{numprint}
\usepackage{nicematrix}
\usepackage[compat=0.6]{yquant}
\usepackage{lipsum}
\usepackage{comment}

\usepackage{subcaption}

\usepackage{pgfplots}
\pgfplotsset{compat = newest}
\usepackage{tikz}
\usetikzlibrary{arrows.meta, shapes.geometric,
                positioning,fit,shadows.blur,
                patterns,calc, matrix,
                tikzmark, backgrounds,
                decorations.pathreplacing}

\usepackage{tikzpagenodes}

\usepackage{algorithm}
\usepackage{algpseudocode}

\usepackage{fp}
\usepackage{xspace}
\usepackage{amsmath,amssymb,amsfonts}
\usepackage{amsthm}
\usepackage{bbm}
\usepackage{quantikz}
\usepackage{placeins}
\usepackage[T1]{fontenc}
\usepackage[colorlinks=true,urlcolor=teal,
            linkcolor=teal,citecolor=teal]{hyperref}
\usepackage[most]{tcolorbox}
\usepackage{listings}
\usepackage{censor}
\usepackage{todonotes}
\usepackage[bold,full]{complexity} 

\usepackage{glossaries-extra}
\setabbreviationstyle[acronym]{long-short}
\glssetcategoryattribute{acronym}{nohyperfirst}{true}

\definecolor{lfdblack}{HTML}{000000}
\definecolor{lfdyellow}{HTML}{E69F00}
\definecolor{lfddgrey}{HTML}{999999}
\definecolor{lfdgreen}{HTML}{009371}
\definecolor{lfdhgrey}{HTML}{beaed4}
\definecolor{lfdred}{HTML}{ed665a}
\definecolor{lfdblue}{HTML}{1f78b4}
\definecolor{bggray}{gray}{0.9}

\newboolean{anonymous}
\setboolean{anonymous}{false}

\addto\extrasenglish{%
}

\newcommand{\reprolink}{https://github.com/ReproPackage/Q25}
\newcommand{\ie}{\emph{i.e.}\xspace}
\newcommand{\eg}{\emph{e.g.}\xspace}
\newcommand{\etal}{\emph{et al.}\xspace}

\makeatletter
\def\calcLength(#1,#2)#3{%
  \pgfpointdiff{\pgfpointanchor{#1}{center}}%
  {\pgfpointanchor{#2}{center}}%
  \pgf@xa=\pgf@x%
  \pgf@ya=\pgf@y%
  \FPeval\@temp@a{\pgfmath@tonumber{\pgf@xa}}%
  \FPeval\@temp@b{\pgfmath@tonumber{\pgf@ya}}%
  \FPeval\@temp@sum{(\@temp@a*\@temp@a+\@temp@b*\@temp@b)}%
  \FProot{\FPMathLen}{\@temp@sum}{2}%
  \FPround\FPMathLen\FPMathLen5\relax
  \global\expandafter\edef\csname #3\endcsname{\FPMathLen}
}
\makeatother

\lstdefinestyle{query}{
  language=SQL,
  stepnumber=1,
  numbersep=10pt,
  tabsize=4,
  showspaces=false,
  showstringspaces=false,
  basicstyle=\linespread{1}\fontfamily{lmtt}\selectfont\small,
  keywordstyle=\color{blue},
  stringstyle=\color{purple},
  upquote=true,
  breaklines=true,
  commentstyle=\color{CadetBlue}
}

\definecolor{mygray}{rgb}{0.643,0.643,0.643}
\newtcolorbox{querybox}[2][]{%
  sidebyside align=top,
  enhanced,
  boxsep=2pt,
  arc=0pt,
  top=-3pt, bottom=-3pt,
  left=2pt, right=0pt,
  colback=white,
  colframe=mygray,
  boxrule=0.5pt,
  leftrule=12pt,
  overlay unbroken and first ={%
      \node[rotate=90,
        minimum width=0.5cm,
        anchor=south,
        yshift=-11pt,
        white]
      at (frame.west) {#2};
    }
}

\newtcolorbox{matrixbox}[2][]{%
  sidebyside align=top,
  enhanced,
  boxsep=0pt,
  arc=0pt,
  left=-1em,
  top=-0.8em,
  boxrule=0pt,
  colframe=bggray,
  colback=bggray,
  leftrule=12pt,
  overlay unbroken and first ={%
      \node[rotate=90,
        minimum width=0.5cm,
        anchor=south west,
        font=\itshape,
        yshift=0pt,
        xshift=0.5em,
        black]
      at (frame.south west) {#2};
    }
}

\DeclareMathOperator*{\argmax}{arg\,max}

\newcommand{\suppweb}{\href{https://github.com/lfd/QSW25-SAT-Strikes-Back}{supplementary website}\xspace}
\newcommand{\repropkg}{\href{https://doi.org/10.5281/zenodo.15464391}{reproduction package}\xspace}

\ifbool{anonymous}{}{\renewcommand{\censor}[1]{#1}}
\ifbool{anonymous}{}{\renewcommand{\blackout}[1]{#1}}
\ifbool{anonymous}{\newcommand{\genemail}[2]{\href{mailto:xxx.xxx@xx.xx}{\blackout{#2}}}}{\newcommand{\genemail}[2]{\href{#1}{#2}}}

\newacronym{qaoa}{QAOA}{Quantum Approximate Optimisation Algorithm}
\newacronym{qubo}{QUBO}{Quadratic Unconstrained Binary Optimisation}
\newacronym{pubo}{PUBO}{Polynomial Unconstrained Binary Optimisation}
\newacronym{wlog}{WLOG}{Without loss of generality}
\newacronym{nisq}{NISQ}{Noisy Intermediate Scale Quantum}
\newacronym{tsp}{TSP}{Travelling Saleseman Problem}
\newacronym{pbf}{PBF}{Pseudo-Boolean Function}
\newacronym{op}{OP}{Optimisation Problem}
\newacronym{vqe}{VQE}{Variational Quantum Eigensolver}
\newacronym{mis}{MIS}{Maximum Independent Set}
\newacronym{sat}{SAT}{Satisfiability}
\newacronym{cnf}{CNF}{Conjunctive Normal Form}
\newacronym{cdcl}{CDCL}{Conflict-Driven Clause Learning}
\newacronym{lsr}{LSR}{Local Structure Reduction}
\newacronym{qecc}{QECC}{Quantum Error Correcting Codes}

\begin{document}
\bstctlcite{BSTcontrol}
\title{SAT Strikes Back: Parameter and Path Relations\\ in Quantum Toolchains}

\author{
\IEEEauthorblockN{\blackout{Lukas Schmidbauer}}
  \IEEEauthorblockA{\blackout{\textit{Technical University of}}\\
    \blackout{\textit{Applied Sciences Regensburg}} \\
    \blackout{Regensburg, Germany} \\
    \genemail{mailto:lukas.schmidbauer@othr.de}{lukas.schmidbauer@othr.de}}
\and
  \IEEEauthorblockN{\blackout{Wolfgang Mauerer}}
  \IEEEauthorblockA{\blackout{\textit{Technical University of}}\\
    \blackout{\textit{Applied Sciences Regensburg}}\\
    \blackout{\textit{Siemens AG, Technology}}\\
    \blackout{Regensburg/Munich, Germany}\\
    \genemail{mailto:wolfgang.mauerer@othr.de}{wolfgang.mauerer@othr.de}}
}

\maketitle

\begin{abstract}
In the foreseeable future, toolchains for quantum computing should offer automatic means of transforming a high level problem formulation down to a hardware executable form.
Thereby, it is crucial to find (multiple) transformation paths that are optimised for (hardware specific) metrics.
We zoom into this pictured tree of transformations by
focussing on $k$-SAT instances as input and their transformation to QUBO, while considering structure and characteristic metrics of input, intermediate and output representations.
Our results can be used to rate valid paths of transformation in advance---also in automated (quantum) toolchains.
We support the automation aspect by considering stability and therefore predictability of free parameters and transformation paths.
Moreover, our findings can be used in the manifesting era of error correction (since considering structure in a high abstraction layer can benefit error correcting codes in layers below).
We also show that current research is closely linked to \emph{quadratisation} techniques and their mathematical foundation.
\end{abstract}


\section{Introduction}
\label{sec:intro}
Although quantum computers promise mathematically backed advantages \cite{Grover_1996}, achieving practical quantum advantage is a herculean task.
On the one hand, hardware has to cope with noise, imperfections, limited amount of inhomogeneous qubits and restricted topology, which narrows down potential applicability---also in the field of error correction.
On the other hand, preprocessing problem formulations has a significant impact on solution quality, performance and deployability onto current hardware \cite{schmidbauer_24_qsw, Schoenberger_2023, Mesman_24,Quetschlich_24,Schnaus_24,Wright_24,Reale_24,Wille_24,yue:2023:qswa,Maniraman_2024}.
However, preprocessing steps (or transformations) occur (in-)between all abstraction layers, starting from an abstract problem formulation down to transpilation onto hardware---leading to mountainous amounts of combinations.
Moreover, transformations also encode NP-complete problems (\eg, mapping and routing problem \cite{Zhang_2021, Cowtan_2019}), which complicates finding optimal solutions.

$k$-\gls{sat} formulations have an enormous field of applications. 
Scheduling problems \cite{akram2024engineeringoptimalparalleltask}, circuit equivalence checking \cite{Kuehlmann2002}, string constraint handling \cite{Vojtech24} and quantum circuit optimisation \cite{shaik_2024optimallayoutsynthesisdeep, Yang24} are some of many applications.
Furthermore, industrial problems induce specific structures and properties into their \gls{sat} formulations \cite{Anstegui2017}. 
It is widely presumed that targeted classical solvers (\eg, \gls{cdcl}) are able to exploit these hidden structures.
Notably, self-similar~\cite{Anstegui2014}, community~\cite{Anstegui2012, Newsham2014} and scale-free structures~\cite{Anstegui2009, Anstegui2009_2} are among relevant properties for industrial \gls{sat} instances.
Take into consideration that different solvers are more or less suited for specific properties of SAT instances.
For example, (satisfiable) random $k$-\gls{sat} instances can be solved using Stochastic Local Search or Look-Ahead solvers \cite{Anstegui2017}.
Transforming a \gls{sat} instance to widely used \gls{qubo} form (\eg, \cite{schoenberger:23:qdsm,Franz_2024}) makes quantum annealing available as a hardware solver\cite{Schoot_2022}. 
Similar to classical solvers, different properties result in varying performance.
For instance, \cite{krueger_2020_SAT} compares randomly generated 3-\gls{sat} instances and their performance on DWave's quantum annealer for two different \gls{qubo} formulations.

The performance of \gls{sat} solvers depends on the size and structure of the instance specific solution space.
This has been widely studied for random $3$-\gls{sat} instances, where the clause-to-variable ratio $\alpha = \frac{m}{n}$ leads to a phase transition at $\alpha \approx \alpha_C = 4.267$ from satisfiable to non-satisfiable instances \cite{Mzard2002}.
For $\alpha \approx \alpha_d = 3.921$, hard instances, inducing metastable states, can be found \cite{Mzard2002}.
For quantum annealing, Gabor \etal~\cite{Gabor2019} show empirically that performance also depends on clause-to-variable ratio $\alpha$ by firstly transforming the input $3$-\gls{sat} representation to a \gls{mis} problem and then to \gls{qubo} (see \autoref{sec:rel_work}).

\begin{figure}[htb]
    \centering
    \includegraphics[]{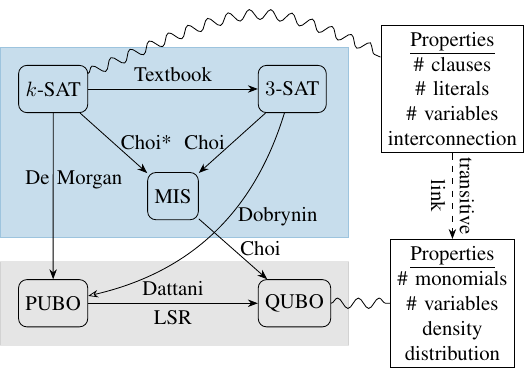}
    \caption{Possible paths from $k$-\gls{sat} to \gls{qubo} via known transformations with special focus on properties induced into resulting \gls{qubo}. Choi* marks a generalised version of the $3$-\gls{sat} to \gls{mis} transformation. 
    Perceived abstraction layer of \emph{NP-complete} problems coloured in \textcolor{lfdblue}{blue} and \emph{Optimisation} problems coloured in \textcolor{lfddgrey}{grey}.}
    \label{fig:Intro_overview}
\end{figure}

From a software engineering perspective, it is therefore essential to know effects of single and transitive transformations on important metrics to get most out of available quantum hardware.
We want to extend above work, by shedding more light on the induced structure and properties, when transforming from $k$-\gls{sat} to \gls{qubo}.
\autoref{fig:Intro_overview} shows a multitude of possible transformations and possible paths from $k$-\gls{sat} to \gls{qubo} alongside their respective properties.
\autoref{sec:fundamentals} introduces fundamental concepts.
They are used in \autoref{sec:rel_work}, which analyses in depth shown transformations from \autoref{fig:Intro_overview}.
\autoref{sec:Experiments} then analyses discussed transformation paths based on their structure and scaling effects on the metrics shown in \autoref{fig:Intro_overview}.
We also discuss implications on quantum software in \autoref{sec:implicationsQS} and conclude in \autoref{sec:ConclAndOutlook}.

The paper is augmented by a comprehensive \repropkg~\cite{mauerer_22_QSaner} and a \suppweb
(links in PDF) that allow for extending our work.

\section{Fundamentals}
\label{sec:fundamentals}
\subsection{SAT}
\gls{sat} is an NP-complete problem that determines if a Boolean formula is satisfiable. 
Although there is a multitude of possible representations of such formulas, \gls{cnf} is a prominently used representation, since every Boolean circuit can be transformed into an equi-satisfiable \gls{cnf} formula in linear time~\cite{Tseitin_1983}.
\glspl{cnf} are also a standard representation for annual \gls{sat} competition~\cite{Froleyks_2021}.

A \gls{sat} formula $\psi(\vec{x})$ in $n$ variables $\vec{x}=(x_1,x_2,\ldots,x_n)$ in \gls{cnf} is a conjunction of $m$ clauses $C_i, i \in \{1,\ldots,m\}$:
\begin{equation}
    \psi(\vec{x}) = \bigwedge_{i=1}^m C_i,\; \vec{x} \in \{0,1\}^n
\end{equation}
A clause $C_i$ consists of a disjunction of positive or negative literals, where a positive literal is a variable $x_i$ and a negative literal is the negation of a variable $\overline{x_i}$.
We denote a positive literal (or variable) by $l$ and negative literal (or negated variable) by $\overline{l}$.
When the number of literals in clauses $C_i, i \in \{1,\ldots,m\}$ is fixed to $k \in \mathbb{N}$, $\psi(\vec{x})$ is an exact-$k$-\gls{sat} instance.
For example,
\begin{equation}
    \psi(\vec{x}) = (x_1 \lor x_2 \lor \overline{x_4} \lor x_5) \land (\overline{x_1} \lor x_3 \lor \overline{x_4} \lor \overline{x_5})
\end{equation}
is an exact $4$-\gls{sat} formula in variables $\vec{x} = (x_1, x_2, x_3, x_4, x_5) \in \{0,1\}^5$.
Hence, there are $2^5$ possible assignments for $\vec{x}$.
If $\exists \vec{x} \in \{0,1\}^n: \psi(\vec{x}) = 1$, we call $\psi(\vec{x})$ satisfiable.
$|C|$ denotes the size of a clause $C$.
The objective of the MAX-$k$-\gls{sat} problem is to find $\vec{x} \in \{0,1\}^n$ such that the number of satisfied clauses is maximised.

\subsection{Maximum Independent Set}
The independent set problem asks for a subset of nodes, such that no node pair in the subset is connected via an edge.
\gls{mis} then asks for the biggest independent set.
\gls{mis} is an NP-hard problem with an NP-complete decision variant (see \gls{sat}). 

\begin{figure}[htb]
    \centering
    \includegraphics[]{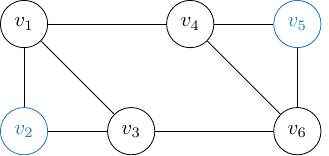}
    \caption{Example of a \gls{mis} $I = \{v_2, v_5\}$ coloured in \textcolor{lfdblue}{blue}. Note that $\{v_1, v_2, v_3\}$ and $\{v_4, v_5, v_6\}$ form a $3$-clique (\ie, a fully connected subgraph).}
    \label{fig:MIS_example}
\end{figure}

More formally, let $G(V,E)$ denote an undirected graph. 
Then, an independent set is a subset of nodes $I \subseteq V$ such that there exists no edge $e = \{i,j\} \in E: i \in I \land j \in I$.
The maximum independent set is (a) not contained in any other independent set (also called maximal condition) and (b) largest with respect to the cardinality of $I$ \cite{Butenko2003, cormen2022introduction}.
\autoref{fig:MIS_example} shows an example graph and a \gls{mis}.

\subsection{PBF, PUBO, QUBO}
A \gls{pbf} is a function
\begin{equation}
    f: \{0,1\}^n \to \mathbb{R}
\end{equation}
that can be uniquely represented by a multi-linear polynomial~\cite{Boros_2002}:
\begin{equation}
  \label{eq:multi_linear_polynomial}
  f(x_1, \ldots , x_n) = \sum_{S\subseteq \{1,\ldots ,n\}} \alpha_S \prod_{j\in S}x_j,
\end{equation}
where $\alpha_S \prod_{j\in S}x_j$ is called a monomial of $f$ and $\alpha_S \in \mathbb{R}$.
The degree (or order) of a monomial is given by $|S|$.

\gls{pubo} refers to the problem of finding $\vec{x} \in \{0,1\}^n$, such that \gls{pbf} $f(\vec{x})$ is maximised or minimised.
\gls{qubo} refers to the same problem, while restricting $f$ to be quadratic or in other words only allowing for monomials of degree at most two. 
For instance, $3 x_1x_2x_3$ is a degree-$3$ monomial, while $\pi x_1x_2$ is a degree-$2$ monomial and thus allowed in \gls{qubo} problems.

We sometimes refer to the \emph{quadratisation} of a higher-order (\ie, degree greater than two) function $f(\vec{x})$ as the reduction of \gls{pubo} to \gls{qubo}.
Technically, a \gls{pbf} $f'(\vec{x},\vec{y})$ is a \emph{quadratisation} of $f(\vec{x})$, if $f'(\vec{x},\vec{y})$ is a quadratic \gls{pbf} ($\operatorname{deg}(f') = 2$) in $\vec{x} = x_1,\ldots ,x_n$ and $\vec{y} = y_{1},\ldots ,y_{m}$, and satisfies~\cite{Boros_2002}:
\begin{equation}
  \label{eq:QuadratizationCriteria}
  f(\vec{x}) = \min_{\vec{y}\in \{0,1\}^m} f'(\vec{x},\vec{y}) \; \forall \vec{x} \in \{0,1\}^n.
\end{equation}
Linked to that is a standard penalty term for iterative \emph{quadratisation} that constrains new variables \cite{Boros_2002}:
\begin{equation}
\label{eq:PenaltyTermIntro}
    \operatorname{p}(x_i,x_j,y_h) = 3y_h + x_ix_j - 2x_iy_h -2x_jy_h.
\end{equation}

\subsection{Graph Representations for SAT and PBF}
To analyse structural properties, we define the following variable incidence graph $G(V,E)$ for \gls{sat} formulas:
Let $\psi(\vec{x}) = \bigwedge_i C_i$ be a \gls{sat} formula in \gls{cnf}. 
Then, $V = \{x_1,\ldots,x_n\}$ is the set of variables. 
Edges are introduced, whenever two variables (negated or not) occur in the same clause $C_i$, that is, edge $e = (x_a, x_b) \in E \Leftrightarrow \exists i: x_a \in C_i \land x_b \in C_i$.
Variable $x \in C$ denotes whether variable $x$ occurs in clause $C$ regardless of negation.

Analogously we define a graph representation for a given \gls{pbf} $f(\vec{x}) = \sum_i \alpha_i M_i$, where $\alpha_i M_i$ enumerates all possible monomials based on $\vec{x}$.
As before nodes $V = \{x_1,\ldots, x_n\}$.
Similarly, edges are introduced, whenever two variables occur in the same monomial $M_i$:
$e = (x_a, x_b) \in E \Leftrightarrow \exists i: x_a \in M_i \land x_b \in M_i \land \alpha_i \neq 0$.

For both graph representations, we disregard self-edges, since in the case of \gls{sat} they would either mean that a variable is redundant in a clause or the clause is trivially satisfiable and in the case of \glspl{pbf}, $x^n = x, n \in \mathbb{N}$.
Take into consideration that the same variable pair can occur in multiple clauses or monomials. 
To keep this information, a multi graph (multiset $E$) could be used.
However, for sake of simplicity, we do not formally introduce a multigraph, but rather denote the number of edges between two nodes by an edge label.

\section{Related Work}
\label{sec:rel_work}
\subsection{Textbook $k$-SAT to $3$-SAT}
\label{ssec:TextbookSAT}
The textbook reduction from $k$-\gls{sat} to $3$-\gls{sat} \cite{cormen2022introduction} provides a simple method to reduce the size of clauses.
Let $t=3$ be the target size.
In essence, we iteratively replace the last $t-1$ variables by a new variable in each clause $C_i$ larger than $t$.
Hence, the size of $C_i$ reduces by $t-2$ and we introduce a new $t$-\gls{sat} clause that contains the negation of the new variable and the formerly replaced ones.
For example, let $\psi(\vec{x}) = (x_1 \lor x_2 \lor \overline{x_3} \lor \overline{x_4} \lor \overline{x_5})$ be a $5$-\gls{sat} instance.
Then,
\begin{equation*}
    \begin{split}
        &(x_1 \lor x_2 \lor \overline{x_3} \lor \overline{x_4} \lor \overline{x_5}) \overset{x_6}{\longrightarrow}\\
        & (x_1 \lor x_2 \lor \overline{x_3} \lor x_6) \land (\overline{x_6} \lor \overline{x_4} \lor \overline{x_5}) \overset{x_7}{\longrightarrow}\\
        & (x_1 \lor x_2 \lor x_7) \land (\overline{x_7} \lor \overline{x_3} \lor x_6) \land (\overline{x_6} \lor \overline{x_4} \lor \overline{x_5}) = \psi'(\vec{x'})
    \end{split}
\end{equation*}
$\psi'(\vec{x'})$ is a $t$-\gls{sat} formula that is satisfiable if and only if $\psi(\vec{x})$ is satisfiable.
Note that this method can easily be extended to reduce to $t$-\gls{sat} with $t>3$ formulas.

\subsection{MAX-k-SAT to QUBO}
Chancellor \etal~\cite{Chancellor2016} provide a \gls{qubo} mapping of MAX-$3$-\gls{sat} and mentions the easily expandability of his construction to MAX-$k$-\gls{sat}\footnote{Technically, Ising Spin Glasses are used, which can easily be mapped to \gls{qubo} by variable substitution.}. 
By counting the number of literals that satisfy a clause for a given assignment via one-hot encoded ancillas, he is able to use penalty terms per clause that match a given MAX-$3$-\gls{sat} problem.
Chancellor \etal also mention that more efficient (in terms of ancilla bits) constructions are possible.
For instance, a $3$-\gls{sat} clause can be implemented into \gls{qubo} by using a single (and not $3$, as before) ancilla bit. 
We demonstrate this mapping, when considering the general mapping of a $k$-\gls{sat} clause to \gls{pubo}, which is then transformed to \gls{qubo} via \textit{quadratisation}.

N\"{u}\ss{}lein \etal~\cite{Nuesslein22} expand the idea of counting the number of literals that satisfy a clause by using binary encoding---lowering the number of ancillas per clause to grow logarithmically in $k$ for $k\geq4$ in a MAX-$k$-\gls{sat} problem.
To be more precise, the number of variables in the resulting \gls{qubo} $N$ is given by:
\begin{equation}
    N = n + m \cdot r(k),  
\end{equation}
where $n$ is the number of variables, $m$ is the number of clauses and 
\begin{equation*}
r(k) =
    \begin{cases}
        0 & \text{if } k=2,\\
        1 & \text{if } k=3,\\
        \lceil \log_2(k+1)\rceil + r(\lceil \log_2(k+1)\rceil) & \text{if } k\geq 4.
    \end{cases}
\end{equation*}
Note that both methods scale linearly in the number of clauses.

\subsection{3-SAT to QUBO}
\label{ssec:3satToQubo}
Recall the definition of \gls{sat} and \gls{mis} from \autoref{sec:fundamentals}.
We now shortly review a known reduction from $3$-\gls{sat} to \gls{mis} (Choi) \cite{choi_2010}. 
Then, we can use a known \gls{qubo} formulation for the \gls{mis} instance.

Let $\psi(\vec{x}) = C_1 \land C_2 \land \ldots \land C_m$ be a $3$-\gls{sat} instance in \gls{cnf}, that is, each clause $C_i$, $i \in \{1,\ldots,m\}$ consists of at maximum $3$ literals (\ie, (negated) variables $x_1,\ldots, x_n$):
$C_i = (l_{i_1} \lor l_{i_2} \lor l_{i_3})$.
For each literal in a clause $C_i$, we create a group of fully connected nodes in graph $G_\text{SAT}(V,E)$ (see \autoref{fig:MIS_example}) \cite{choi_2010}.
Additionally, every conflicting literal pair (\ie, $l_{i_s} = \overline{l_{j_t}}$; $i \neq j$) in $G_\text{SAT}$ introduces an edge $(i_s, j_t)$.
Then, the following statements are logically equivalent:
\begin{itemize}
    \item $\psi(\vec{x})$ is satisfiable.
    \item $G_\text{SAT}$ has a \gls{mis} of size $m$.
\end{itemize}
The known \gls{qubo} formulation of a \gls{mis} \cite{choi_2010} incorporates its graph $G$ and was recently used by Zielinski \etal~\cite{Zielinski_2023} to compare it to other transformations to \gls{qubo}.
Its optimisation function is given by \cite{choi_2008}:
\begin{equation}
    \gamma(x_1, \ldots, x_n) = \sum_{i \in V} c_ix_i - \sum_{(i,j) \in E} J_{ij}x_ix_j,
\end{equation}
where $c_i = 1$ is the weight for nodes in the graph\footnote{\gls{sat} reduces to an unweighted graph.} and $J_{ij} > 1 \; \forall (i,j) \in E$.
Maximising function $\gamma(x_1, \ldots, x_n)$ solves the \gls{mis} problem.
In particular, the set of nodes $\text{mis}(G) = \{i \in V: x_i^* = 1\}$, where $(x_1^*, \ldots, x_n^*) = \argmax_{(x_1, \ldots, x_n) \in \{0,1\}^n} \gamma(x_1, \ldots, x_n)$, corresponds to \gls{mis}.

Zielinski \etal~\cite{Zielinski_2023} structure $3$-\gls{sat} clauses into four types, depending on (negated) literals.
They present an algorithm to search for valid \gls{qubo} representations\footnote{Note that there are infinitely many valid representations.} $Q_i, i \in \{1,\ldots,m\}$ for single $3$-\gls{sat} clauses and then combine each $Q_i$ to a \gls{qubo} $Q$, using methods presented in \cite{Chancellor2016}.
This method results in a \gls{qubo} size of $n + m$, where $n$ is the number of variables in the original \gls{sat} instance and $m$ is the number of clauses.
Note that defining types of $k$-\gls{sat} clauses, as before for $3$-\gls{sat}, scales exponentially in $k$.
However, textbook reductions from $k$-\gls{sat} to $3$-\gls{sat} can be applied prior to type definition (see \autoref{sec:fundamentals}).

This method is closely linked to \emph{quadratisation} techniques found in the works of Boros \etal~\cite{Boros_2002, Boros_2014, Boros_2019}:
We can reformulate the above stated as firstly defining a family of valid \emph{quadratisations} for clause types that use a single ancilla variable.
Alternatively, one can use the methods presented in \autoref{ssec:rel_work_kSatToPubo} to formulate higher-order \glspl{pbf} $f_i, i \in \{1,\ldots,m\}$ and then apply one of many reductions from \cite{Dattani_2019} to arrive at a similar quadratic function for clauses. 
Secondly, a \gls{pbf} for the original $3$-\gls{sat} instance is determined by summation:
\begin{equation}
    f = \sum_{i = 1}^m f_i.
\end{equation}
Dobrynin \etal~\cite{Dobrynin_2024} point to a similar argument by incorporating different penalty terms in their \gls{pubo} to \gls{qubo} transformation.
In \cite{Nuesslein2023nmApproach}, the $n + m$ approach provides different penalty terms with fewer interactions than the standard penalty (see \autoref{sec:fundamentals}). 
Despite these penalties not obeying the \emph{quadratisation} criteria, they preserve at least one minimum\footnote{We refer to the penalties given for $(a\lor b \lor \overline{c})$ and $ (a\lor \overline{b}\lor \overline{c})$.}.
Although, every isolated $k$-\gls{sat} clause $C_i$ has $2^{k} -1$ satisfying variable assignments,
penalty terms that do not obey the \emph{quadratisation} criteria potentially reduce the set of valid possible solutions for isolated clauses.
However, since clauses $C_i, i \in \{1,\ldots,m\}$ typically share variables or literals in a given $k$-\gls{sat} formula, it is important to preserve all local satisfying solutions for clauses.
If not, satisfying variable assignments for the given $k$-\gls{sat} formula do not map to minimum in \gls{qubo}.
Take into consideration that the effect of this local inaccuracy is especially pronounced on $k$-\gls{sat} formulas with high inter-clause connectivity, high number of clauses or high $k$.
We therefore emphasize the importance to adhere to the \emph{quadratisation} criteria (see \autoref{eq:QuadratizationCriteria}), when building upon small solution sets. 

\subsection{$k$-SAT to MIS to QUBO}
Through a similar argument for the reduction from $3$-\gls{sat} to \gls{mis} (see \autoref{ssec:3satToQubo}), we extend this study by comparing \gls{qubo} mappings for $k$-\gls{sat} instances.
Analogously, each exact $k$-\gls{sat} clause $C_i$ creates a fully connected sub-graph in $G$ with $k$ nodes.
As before, conflicting literals in different sub-graphs are connected. 
Note that if there are conflicting literals in the same sub-graph, they are already connected by construction.
Hence, for exact $k$-\gls{sat} with $m$ clauses, there are $k \cdot m$ nodes in the graph.

As a side note, the size of resulting \gls{qubo} is at least twice the number of variables squared, when all literals are used in the SAT formula. 
As with the $3$-\gls{sat} to \gls{mis} construction, choosing $J_{ij} > min(c_i, c_j)$, $c_i > 0$ is required.

\subsection{$k$-SAT to PUBO}
\label{ssec:rel_work_kSatToPubo}
Let $\psi(\vec{x})$ be an exact $k$-\gls{sat} formula in \gls{cnf} with clauses $C_i$, $i \in \{1, \ldots, m\}$.
For example, let $C_1 = (\overline{x_1} \lor \overline{x_2} \lor \overline{x_3})$ and $C_2 = (x_2 \lor x_3 \lor \overline{x_4})$. 
Then, contrary to Choi's reduction, we can encode a negated variable $x_i$ as $1 - x_i$.
The same applies for clause negation, by using DeMorgan's rules for multiple variables (see also \cite{Dobrynin_2024}).
Hence, equivalent \glspl{pbf} for clauses $C_1$ and $C_2$ are given by
\begin{equation}
\label{eq:PBFforClauseExample}
    \begin{split}
        f_{C_1}(x_1,x_2,x_3) &= 1 - x_1x_2x_3\\
        f_{C_2}(x_2,x_3,x_4) & = 1 - (1-x_2)(1-x_3)x_4.
    \end{split}
\end{equation}
Their sum then encodes the decision problem: 
\begin{equation*}
    f_\text{SAT}(x_1, x_2, x_3, x_4) = f_{C_1}(x_1,x_2,x_3) + f_{C_2}(x_2,x_3,x_4).
\end{equation*}
A similar construction for $3$-\gls{sat} formulas can be found in \cite{Hizzani_2023}.
Note that if a clause $C_i$ evaluates to true the corresponding \gls{pbf} $f_{C_i}$ evaluates to $1$. 
If $C_i$ evaluates to false, then $f_{C_i}$ evaluates to $0$---effectively encoding $C_i$ into a maximisation problem.
The minimisation problem can easily be obtained by negating $f_{C_i}$.
For both maximisation and minimisation problem, the \gls{sat} formula is satisfiable, iff 
\begin{equation}
    \exists \vec{x} \in \{0,1\}^n: |f_\text{SAT}(\vec{x})| = m.
\end{equation}
If $|f_\text{SAT}(\vec{x})| = s$, then there are $s$ many satisfied clauses for the assignment $\vec{x}$, which can be used for the generalised optimisation of MAX-\gls{sat}.

In general, for exact $k$-\gls{sat} formulas, this leads to monomials of degree $k$ in the resulting \gls{pbf}.
Monomials of degree smaller than $k$ occur whenever non-negated variables appear in a clause $C_i$.
Note that if a clause $C_i$ consists of $k$ negated variables (\eg, $(\overline{x_1} \lor \overline{x_2} \lor \overline{x_3} \lor \overline{x_4}\})$), then the resulting \gls{pbf} has a single degree-$k$ monomial (\eg, $x_1x_2x_3x_4$). 
Monomials can can be \emph{quadratised} via either $\lceil log_2(k)\rceil -1$ or a single extra variable---depending on whether $\alpha_S$ is positive or negative~\cite{Boros_2019, Dattani_2019}.
Although redundant clauses are excluded from the input \gls{sat} formula, the resulting degree-$k$ monomial can occur in other transformed \glspl{pbf} $f_{C_j}$.
For example, let $C_j = (x_1 \lor \overline{x_2} \lor \overline{x_3} \lor \overline{x_4}\})$. 
Then $x_1x_2x_3x_4$ occurs in $f_{C_j}(\vec{x}) = 1 - (1-x_1)x_2x_3x_4$ \footnote{We do not consider prefactor $\alpha$ here.}, which offers potential to efficiently \emph{quadratise} multiple occurrences at once.

Conversely, assume a clause $C_i$ consists of $k$ distinct and non-negated variables (\ie, no negative literal). 
Then, the resulting \gls{pbf} $f_{C_i}$ has all possible monomials of degree $i$, $i \in \{1, \ldots, k\}$.
Since there are $\binom{k}{i}$ many possible degree-$i$ monomials, clause $C_i$ introduces 
\begin{equation*}
    \sum_{i = 0}^{k} \binom{k}{i} = 2^k - 1
\end{equation*}
many monomials.
If one would brute force reduce all degree-$i$ ($i > 2$) monomials via a single variable in every clause\footnote{This term is a lower bound, since $\lceil log_2(k)\rceil -1$ extra variables are required for positive degree-$k$ monomials \cite{Boros_2019}.}, there would be
\begin{equation*}
    m \cdot \sum_{i = 3}^{k} \binom{k}{i} = m \cdot \left(2^k - \binom{k}{2} - k - 1\right)
\end{equation*}
many new variables in $f_\text{SAT}$.
However, arbitrary \gls{sat} instances usually do not fit into one of the above categories (all positive or all negative literals) and hence a reduction method exploiting the inner structure of $f_\text{SAT}$ is needed.

\subsection{PUBO to QUBO}
\label{ssec:PUBOtoQUBO}
\begin{figure}[htb]
    \centering
    \includegraphics[]{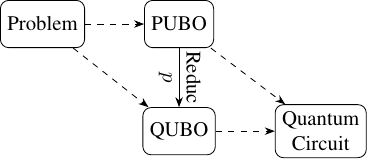}
    \caption{Abstracted \gls{pubo} / \gls{qubo} relation. (Dashed) lines: (Multiple intermediate) transformation(s).}
    \label{fig:rel_puboqubo}
\end{figure}
Let \gls{pubo} and \gls{qubo} denote two formulations for the same problem (see \autoref{fig:rel_puboqubo}).
It is possible to encode both \gls{pubo} and \gls{qubo} into quantum circuits and transpile them onto hardware that features at most two qubit operations.
On the one hand the \gls{pubo} formulation leads to necessary decompositions of higher-order gates. 
On the other hand, the \gls{qubo} formulation has to cope with encoding the same information and therefore has potentially more gates than the original \gls{pubo}.
Reducing the degree of monomials in a \gls{pubo} $f$ can benefit circuit metrics (\ie, circuit depth, distribution of gates and number of gates), when, for example, creating a \gls{qaoa} circuit \cite{schmidbauer_24_qsw}. 
Note that these results stem from an empirical study, featuring an industry relevant Job-Shop Scheduling problem with at most degree-$4$ monomials. 
Contrary to that, Campbell and Dahl \cite{Campbell_2021} execute small instances of the four corner graph colouring problem with \gls{qaoa} and COBYLA and found better results when directly using their \gls{pubo} formulation.

There are many possible transformations from \gls{pubo} $f_P$ to \gls{qubo} $f_Q$---including monomial-wise reductions or considering specific monomial properties \cite{Dattani_2019}. 
When structural properties cannot be leveraged to simplify $f_P$, new variables are introduced to reduce the degree of $f_P$. 
A fast \textit{quadratisation} algorithm (\gls{lsr}) was published in \cite{schmidbauer:24:fastQuadratisation}, which uses iterative \textit{quadratisation} and has a free parameter $p \in [0,1]$.
Parameter $p$ influences properties of the resulting \gls{qubo} $f_Q$. 
For instance, a higher value of $p$ can lead to less variables, but higher degree-$2$ density in $f_Q$ and vice versa.

Mapping a given logical quantum circuit onto hardware (that depends on given \gls{qubo}) often requires the use of SWAP-gates to accommodate for missing connections. 
Finding balance between connectivity in \gls{qubo} and number of qubits is performance-relevant and thus should be considered by an automated (quantum) toolchain.
Percentile $p$ is a simple tunable parameter for this task.

\section{Experiments}
\label{sec:Experiments}
Recall that introduced MAX-$k$-\gls{sat} to QUBO formulations, as well as Choi's reduction via \gls{mis} scale with the number of clauses, but do not consider inter-clause relations (\ie, the inner structure).
Contrary, the iterative reduction is able to exploit the inner structure of the resulting \gls{pubo} formulation, when \emph{quadratising} to \gls{qubo}---not only requiring less extra variables, but also influencing structure of resulting \gls{qubo}. 
Hence, comparing these methods is of interest for the following experiments.
Our experiments also extend depicted research by considering structural properties of input, intermediate and output problem representations. 
\subsection{Setup}
Let $V$ be the set of variables.
Then $L = V \cup \{\overline{x}: x \in V\}$ is the set of literals, where $\overline{x}$ is the negation of $x$.
For all experiments, we randomly sample $k$ literals from $L$ and repeat that step $m$ times (number of clauses).
Since the number of actually used variables in $\psi(\vec{x})$ can be less than $|V|$, we use the number of actually used variables as the x-axis of our graphs.
At first we shed light on the scaling effects and then analyse the structural properties of (intermediate) representations for a single instance.
Take into consideration that, for $k=3$, textbook reduction has no effect on its input.
Hence, Choi and Choi*, as well as Dobry. and DeMorg. perform similarly among all figures for $k=3$.

\subsection{Results}
\begin{figure*}[!htb]
    \centering
    \includegraphics[]{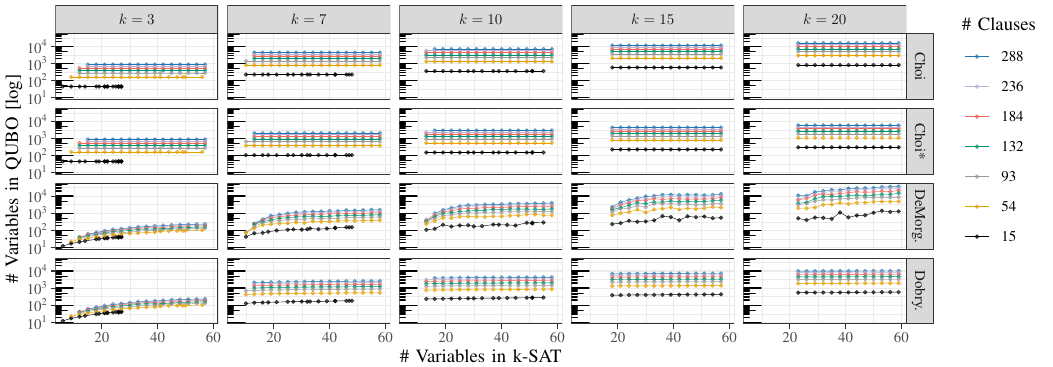}
    \caption{Number of variables in $k$-\gls{sat} (x-axis) vs number of variables in resulting \gls{qubo} (y-axis), coloured by the number of clauses. Horizontal facet: Transformation path. Vertical facet: $k$ in $k$-\gls{sat}. For DeMorg. and Dobry., $p=1$.} 
    \label{fig:Res_variables}
\end{figure*}

\begin{figure*}[!htb]
    \centering
    \includegraphics[]{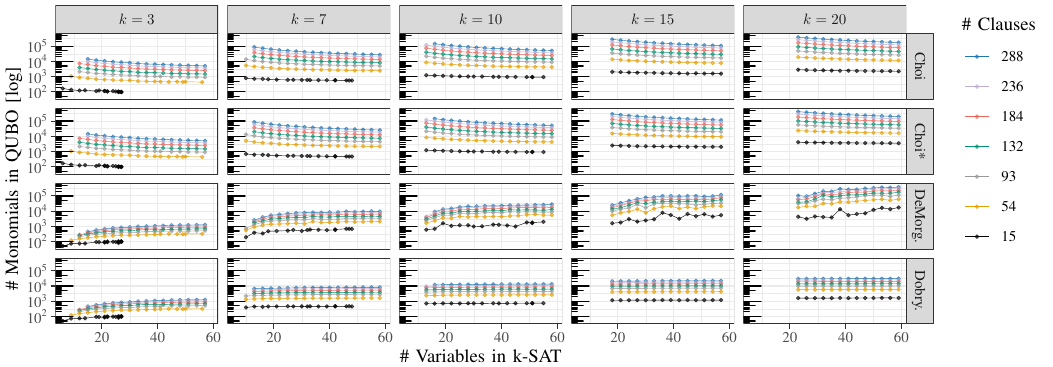}
    \caption{Number of variables in $k$-\gls{sat} (x-axis) vs number of monomials in resulting \gls{qubo} (y-axis), coloured by the number of clauses. Horizontal facet: Transformation path. Vertical facet: $k$ in $k$-\gls{sat}. For DeMorg. and Dobry., $p=1$.}
    \label{fig:Res_variablesMonomials}
\end{figure*}

\begin{figure*}[!htb]
    \centering
    \includegraphics[]{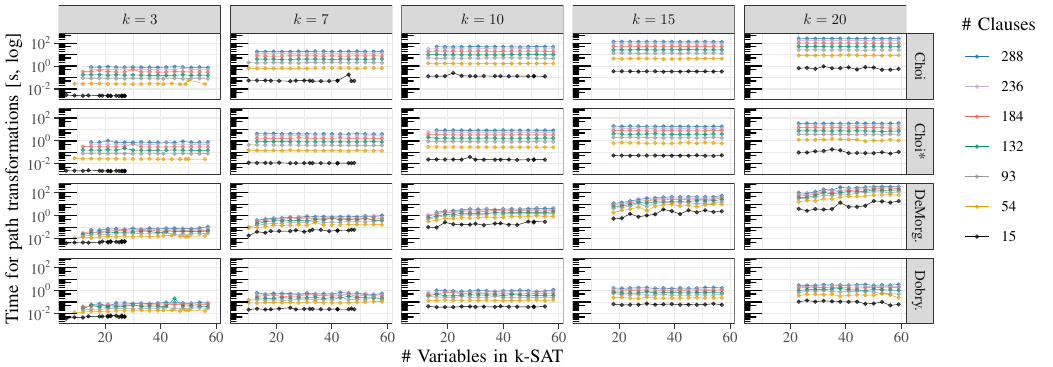}
    \caption{Number of variables in $k$-\gls{sat} (x-axis) vs transformation time (y-axis), coloured by the number of clauses. Horizontal facet: Transformation path. Vertical facet: $k$ in $k$-\gls{sat}. For DeMorg. and Dobry., $p=1$.}
    \label{fig:Res_variablesTime}
\end{figure*}

\begin{figure*}[!tb]
    \centering
    \includegraphics[width=\textwidth]{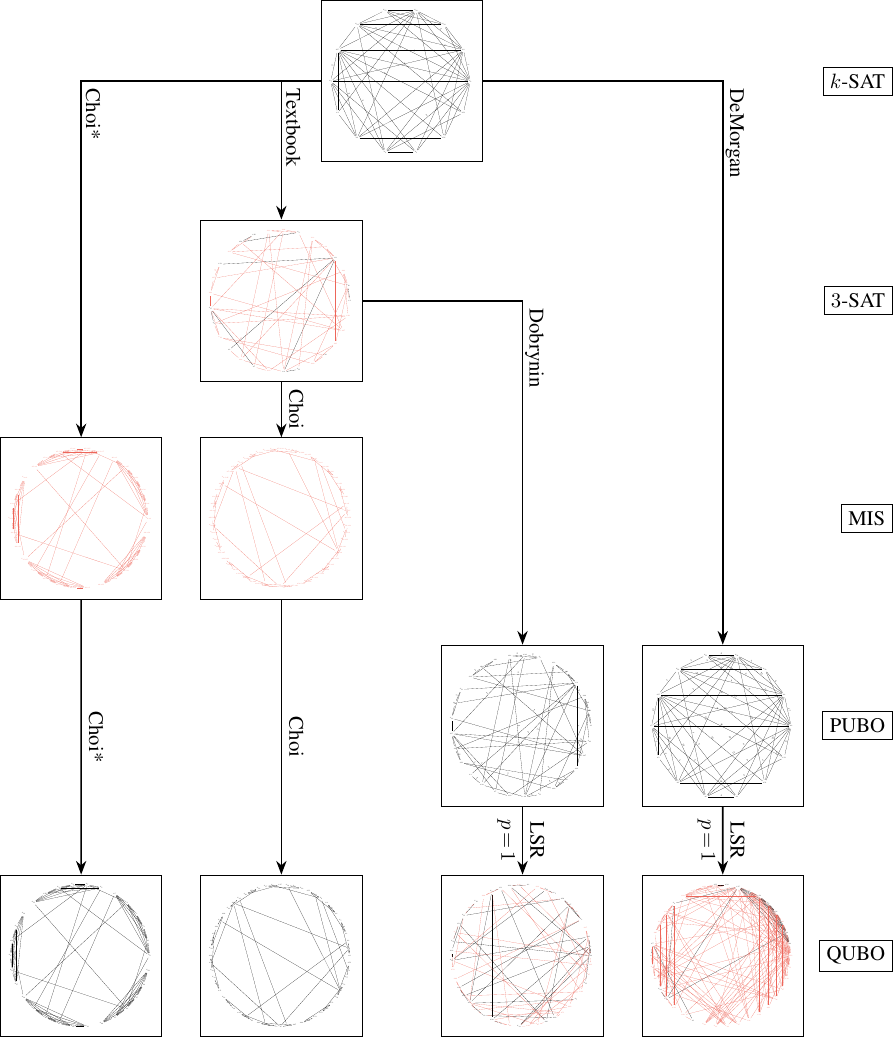}
    \caption{Structural graph evolution of transformations and transformation paths, starting at a $6$-\gls{sat} instance with $5$ clauses and $14$ variables. Choi* marks a generalised version of Choi's $3$-\gls{sat} to \gls{mis} to \gls{qubo} transformation. 
    Nodes are placed, according to the largest clique they occur in---leading to clustered nodes. 
    Newly introduced nodes and edges are shown in \textcolor{lfdred}{red} compared to the last representation. Due to their unique construction, \gls{mis} graphs do not adopt edges or nodes.}
    \label{fig:BigGraphPlot}
\end{figure*}

\autoref{fig:Res_variables}, \ref{fig:Res_variablesMonomials} and \ref{fig:Res_variablesTime} have equal structure: The x-axis represents the number of variables in the input $k$-\gls{sat} instance.
Each horizontal facet shows one of the tested variants, that is, Choi (Textbook reduction to $3$-\gls{sat}, followed by \gls{mis} to \gls{qubo}), Choi* (generalised version, \ie, direct \gls{mis} to \gls{qubo}), Dobry. (Textbook reduction to $3$-\gls{sat}, followed by direct \gls{pubo} mapping and \emph{quadratisation}) and DeMorg. (direct \gls{pubo} mapping, followed by \emph{quadratisation}).
\autoref{fig:BigGraphPlot} also gives a bird eye view on their relation.
Each vertical facet represents $k$ in the input instance, while the colour indicates the number of clauses $m$ in the input instance.

\autoref{fig:Res_variables} shows the number of variables in resulting \gls{qubo} on its y-axis (log scale). 
For each method, the number of clauses increases the number of variables.
For both Choi and Choi*, the increase is linear in the total size of clauses, with respect to the input $3$- or $k$-\gls{sat} instance.
However, Choi has to cope with increased total clause size due to introducing extra variables (see \autoref{ssec:TextbookSAT}) and hence, scales worse than $k \cdot m$.
In particular, the resulting $3$-\gls{sat} formula has $(k-2)\cdot m$ clauses and therefore a total size of $3(k-2)\cdot m$ \footnote{Note that $3(k-2)\cdot m > k \cdot m \Leftrightarrow k > 3 \; \forall m \in \mathbb{N}$.}.
The effects of the textbook reduction are also visible in Dobry., which leads to small dependence on the number of variables in $k$-\gls{sat}, since variables in $3$-\gls{sat} clauses after textbook reduction contain at least $1/3$ new (negated) variables (depending on the particular pair choice in iterative textbook reduction).
Conversely, DeMorg. scales with the number of variables in $k$-\gls{sat}, but introduces less variables up to $k=7$ compared to Choi* and Dobry. and up to $k=10$ for Choi and Dobry..

\autoref{fig:Res_variablesMonomials} shows the number of monomials (mostly interactions) in \gls{qubo} on its y-axis (log scale).
DeMorg. and Dobry. scale similarly with according arguments as before for \autoref{fig:Res_variables}.
For Choi and Choi*, the number of monomials decreases as the number of variables in $k$-\gls{sat} increases.
Recall that their construction (see \autoref{sec:rel_work}) introduces interconnections in \gls{mis} graph, whenever there are conflicting literals in different clauses.
Hence, increasing the number of variables in randomly sampled $k$-\gls{sat} instances, decreases the expected number of conflicts.
Interestingly, this effect also transitively applies through textbook reduction to Choi, since the size of cliques reduces, although the total size of $3$-\gls{sat} clauses increases compared to $k$-\gls{sat}. 

\autoref{fig:Res_variablesTime} shows the total time for the $k$-\gls{sat} to \gls{qubo} reduction in seconds on its y-axis (log scale), which is relevant for time constraint applications.
It is evident that the runtime mostly depends on the number of clauses and $k$ for Choi, Choi* and Dobry.---with Dobry. featuring consistently lower values than Choi, albeit both using textbook reduction.
Up to $k=20$, generalised Choi* is faster than Choi, although it has to generate $20$-cliques in \gls{mis} graph.
In other words, the number of edges in the \gls{mis} graph per $k$-\gls{sat} clause is quadratic in $k$, whereas the textbook reduction introduces $k-2$ $3$-\gls{sat} clauses per $k$-\gls{sat} clause.
Take into consideration that directly mapping $k$-\gls{sat} to \gls{pubo} introduces an exponential amount of monomials in the number of positive literals, due to term expansion. 
We therefore do not recommend this method beyond a certain $k_\alpha > 20$.
However, DeMorg. outperforms Choi and Choi* up to $k=10$ and more than $132$ clauses.
Note however that the underlying implementation can be further optimised in time.

\begin{figure*}[!htb]
    \centering
    \includegraphics[]{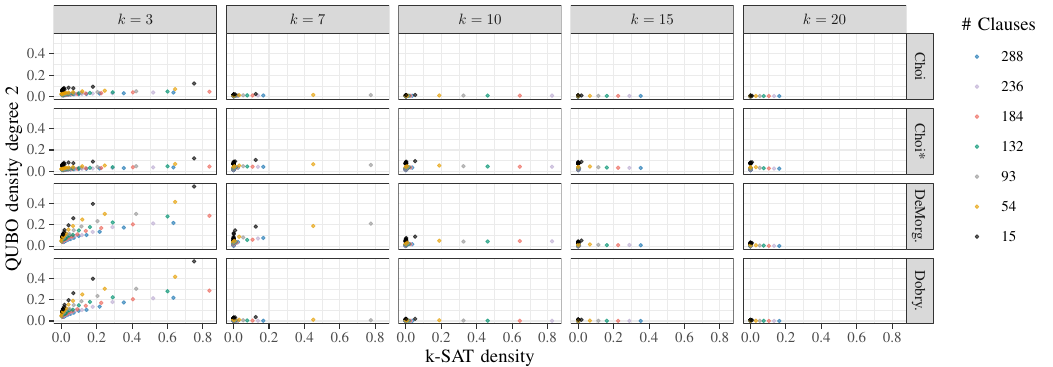}
    \caption{$k$-\gls{sat} density (\ie, ratio of actual to possible clauses; x-axis) vs \gls{qubo} density for degree-$2$ monomials (\ie, ratio of actual to possible degree-$2$ monomials) (y-axis), coloured by the number of clauses. Horizontal facet: Transformation path. Vertical facet: Value of $k$ in $k$-\gls{sat}. For DeMorg. and Dobry., $p=1$.}
    \label{fig:Res_density}
\end{figure*}

\autoref{fig:Res_density} shows the ratio of actual to possible clauses and degree-$2$ monomials on the x- and y-axis respectively.
Each method tends to generate sparse \gls{qubo} representations for $k\geq 10$.
For $k = 3$, DeMorg. and Dobry. result in significantly denser \glspl{qubo} than Choi and Choi*, since they do not introduce additional variables.
For $k = 10$, Choi* and DeMorg. provide higher density due to less introduced variables, as discussed for \autoref{fig:Res_variables}.
Also note that the randomly sampled input $k$-\gls{sat} instances become less dense as $k$ increases.

\begin{figure*}[!htb]
    \centering
    \includegraphics[]{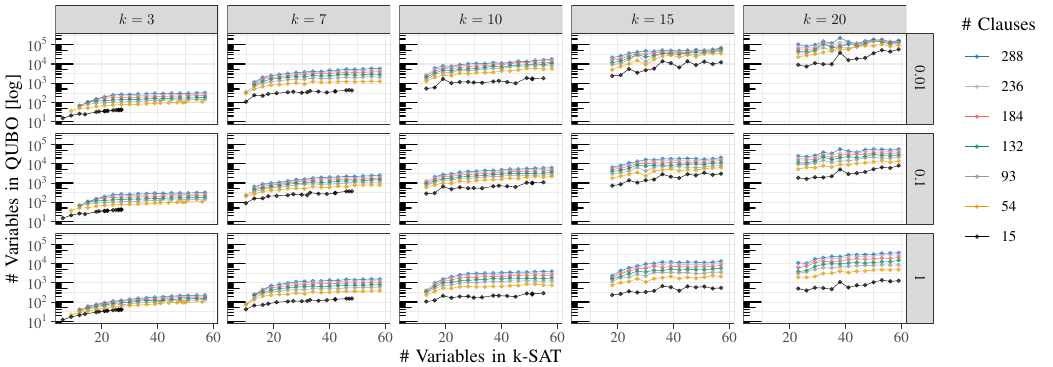}
    \caption{Number of variables in $k$-\gls{sat} (x-axis) vs number of variables in resulting \gls{qubo} (y-axis), coloured by the number of clauses. Horizontal facet: Percentile $p$ for direct mapping to \gls{pubo} (DeMorgan). Vertical facet: $k$ in $k$-\gls{sat}.}
    \label{fig:Res_variablesPercentile}
\end{figure*}

\autoref{fig:Res_variablesPercentile} shows the number of variables in \gls{qubo} on its y-axis (log scale). 
Contrary to previous figures, where $p=1$, its horizontal facets show different percentiles $p$ (see \autoref{ssec:PUBOtoQUBO}) for \emph{quadratisation} and method DeMorg..
In general, higher percentile $p \in [0,1]$ leads to less variables in \gls{qubo}, since higher $p$ leads to the iterative reduction of variable pairs that occur in more monomials per iteration. 
Note however that finding the minimum number of new variables for a valid \emph{quadratisation} is NP-hard \cite{Boros_2002}---indicating potential instabilities in the heuristic polynomial time approach. 
Be that as it may, \autoref{fig:Res_variablesPercentile} suggests a relatively stable process for the tested input instances---supporting the general statement for $p$.


Apart from scaling of metrics, the internal structure and the effect of transformations in each step are relevant to extrapolate our findings to other input problems.
\autoref{fig:BigGraphPlot} shows this exemplary for a single $6$-\gls{sat} instance with 5 clauses and 14 variables.
Each path of transformations has intermediate representations, which are shown as graphs, introduced in \autoref{sec:fundamentals}.
Note that comparing graphs visually can be deceiving, when using arbitrary node positions and ordering.
Therefore, only for the visualisation in \autoref{fig:BigGraphPlot}, we sort nodes according to the highest clique they occur in (see \autoref{alg:node_sort}), where \textsc{Max\_Clique($G$)} returns a list of nodes, contained in a maximum clique of $G$ and operation ``$+$'' (\autoref{alg:node_sort}: l. 3) appends the set of nodes $C$ to list $V'$.
\begin{algorithm}
	\caption{Sorted nodes by clique size.}\label{alg:node_sort}
	\hspace*{\algorithmicindent} \textbf{Input} Graph $G(V,E)$ \hfill\phantom{x}\\
	\hspace*{\algorithmicindent} \textbf{Output} Sorted list of nodes $V'$ \hfill\phantom{x}
	\begin{algorithmic}[1]
		\While{$V \neq \emptyset$}
		      \State $C \gets \Call{Max\_Clique}{G}$
            \State $V' \gets V' + C;\; V \gets V \setminus C$
		\EndWhile
		\State return $V'$
	\end{algorithmic}
\end{algorithm}
In addition to node positioning, we show the difference in nodes and edges to the previous representation by colouring new nodes and edges in \textcolor{lfdred}{red}.
Also recall that edge labels correspond to multiplicity (see \autoref{sec:fundamentals}).
Hence, an edge $e$ with label $3$ between nodes $i$ and $j$ corresponds to $3$ edges.
Therefore, pair $\{i,j\}$ occurs in $3$ clauses (in case of \gls{sat}) or $3$ monomials (in case of \gls{pbf}).
\autoref{fig:BigGraphPlot} shows a subtle characteristic for Choi's reductions, namely $k$-cliques, where $k$ depends on the input \gls{sat} instance.
Hence, shown \gls{mis} graph of Choi* has $5$ $6$-cliques, since there are $5$ clauses and the input is a $6$-\gls{sat} instance.
Likewise, shown \gls{mis} graph of Choi has $20$ $3$-cliques, since there are $20$ clauses in the input $3$-\gls{sat} instance.
Note that the size of cliques is upper bounded by $k$, since a variable in clause $C_i$ cannot be in conflict with all variables of another clause $C_j$---due to the construction of $k$-\gls{sat} instances.
Since we exclude self-edges or linear terms in \gls{pbf} (see \autoref{sec:fundamentals}), the resulting \gls{qubo} graph for Choi(*) is isomorph to the \gls{mis} graph.

This is in contrast to the methods that map a $k$-\gls{sat} instance directly to \gls{pubo}---leading to monomials of degree less than or equal to $k$.
Since, in general, the \emph{quadratisation} of \gls{pubo} requires additional variables and constraint terms (see \autoref{ssec:PUBOtoQUBO}), resulting \glspl{qubo} change in structure and variables.
Note that each \gls{pubo} graph is isomorphic to its $3$- or $k$-\gls{sat} graph up to edge labels, which increase.
Increasing edge labels are a result of degree-$t, t<k$ monomials.
After \emph{quadratisation}, each \gls{qubo} graph has edge label $1$, which is not shown explicitly.
\autoref{fig:BigGraphPlot} shows many more \textcolor{lfdred}{red} edges in \gls{qubo} in the direct mapping (DeMorgan) compared to firstly reducing to $3$-\gls{sat} (Dobrynin). 
\textcolor{lfdred}{Red} edges in \gls{qubo} are a result of the quadratic penalty term (see \autoref{eq:PenaltyTermIntro}) and quadratic monomials.
For example, term $x_1x_2x_3x_4 \overset{x_1x_2=y_1}{\longrightarrow} y_1x_3x_4 + p(x_1,x_2,y_1) \overset{x_3x_4=y_2}{\longrightarrow} y_1y_2 + p(x_1,x_2,y_1) + p(x_3, x_4, y_2)$ leads to a new edge $\{y_1,y_2\}$ and at least $2$ more edges per penalty term.

In summary, \emph{quadratisation} spreads the information encoded by \gls{pubo} over new variables. 
However, directly mapping $k$-\gls{sat} to \gls{pubo} leads to more concentrated regions of high and low connectivity, while first mapping to $3$-\gls{sat} leads to a more uniform distribution of graph connectivity.
The varying spread of interactions is also visible for Choi's reductions.
One can take advantage of these variations in later steps, when, for example, transpiling a circuit onto specific hardware topologies.

\subsection{Qualitative Extrapolation}
Motivated by structure inducing industry relevant problems and based on discussed results for random exact $k$-\gls{sat} instances, we make qualitative arguments for different input $k$-\gls{sat} formulas with varying structure.
This enables estimating structural properties of resulting \gls{qubo} for structured or degenerate input instances.
We therefore consider:
\begin{itemize}
    \item[] (1) High degeneracy in $k$
    \item[] (2) High inter-clause connections (communities/cliques)
\end{itemize}
In the following, we assume that the input is no longer necessarily given as an exact $k$-\gls{sat} instance.
Therefore, clauses $C_i, i \in \{1,\ldots,m\}$ can have different size---up to $k$.
Let $\psi(\vec{x}) = \bigwedge_i C_i$ be a \gls{sat} formula in \gls{cnf}.

\paragraph{Textbook $k$-SAT to $3$-SAT}
The transformation from $k$-\gls{sat} to $3$-\gls{sat} solely depends on the number of clauses of size $k>3$.
Therefore, inter-clause relations do not affect this transformation (2). 
Note that due to newly introduced variables, there cannot be redundant clauses.
A high degeneracy in $k$ or less clauses of size $k$ lead to less new variables (1).
To be more precise, for a clause of size $k$, $k-3$ new variables are introduced.
Take into consideration that there are variations of the introduced textbook reduction (see \autoref{sec:fundamentals}) that, for instance, choose different variable pairs in the iterative process. 
This affects the inner structure and interconnectivity of resulting $3$-\gls{sat} clauses.

\paragraph{Choi(*)}
Recall that in a first step Choi's reduction creates fully connected sub-graphs for literals in each clause. 
Hence, a high degeneracy in $k$ (1) directly influences size of cliques in \gls{mis} graph.
Since resulting \gls{qubo} graph is isomorphic to \gls{mis} graph, it is also directly influenced.
Since Choi's reduction introduces edges, whenever conflicting literals occur in different clauses, a high conflicting inter-clause connectivity (2) leads to more densely connected cliques in \gls{mis} graph.
Take into consideration that influence from textbook reduction are transitively applied, when firstly reducing to $3$-\gls{sat}.

\paragraph{SAT to PUBO}
Directly mapping \gls{sat} to \gls{pubo}, as shown in \autoref{sec:rel_work}, is locally dependent on clauses and their structure. 
For the number of introduced monomials, differentiating between positive and negative variables is decisive, as the number of monomials scales exponentially with the number of positive variables in clauses. 
Hence, a high degeneracy in $k$ (1) or reducing the size of clauses results in less monomials in the resulting \gls{pubo}. 
The number of variables in \gls{pubo} is independent of structural properties in $\psi(\vec{x})$.
Compared to random instances, a high inter-clause connection (2) potentially increases chances that a variable pair occurs in multiple monomials---enabling to reduce the same variable pair in multiple monomials at once and thereby decreasing the number of introduced variables when transforming to \gls{qubo}.
As before, transitive effects from textbook reduction apply---potentially decreasing inter-clause connectivity.

\section{Implications on Quantum Software}
\label{sec:implicationsQS}
At a certain $k_\alpha$ it becomes infeasible to encode $k_\alpha$-\gls{sat} clauses directly into \gls{pubo} form. 
Although we demonstrate scaling behaviour in time and size for $k$, the concrete value for $k_\alpha$ depends on a concrete application.
An upstream reduction of clause size by, for example, textbook reduction (see \autoref{sec:fundamentals}), makes \gls{pubo} a well suited abstraction for quantum toolchains---also considering their expressivity in other problem formulations \cite{schmidbauer_24_qsw} and potential speedup over \gls{qubo} in annealing \cite{nagies2025}.
Choi(*)'s reduced \gls{qubo} has high connectivity among former clauses in $k$-\gls{sat}.
It can also be combined with textbook reductions, to increase the number of variables, but decrease clique-size and keep its regular structure.
Ultimately a hardware executable representation is needed and hence further transformation steps. 
A quantum toolchain can leverage these results by choosing suitable paths and parameters based on later steps. 
For example, a quantum hardware that features high local, but low global connectivity, benefits from the structure of Choi(*)'s reduction, whereas a low qubit count hardware benefits from \gls{pubo} to \gls{qubo} reduction in the case of $3$-\gls{sat} and parameter $p=1$.

Higher order formulations (\ie, increasing $k$) of $k$-\gls{sat}, tend towards sparsely connected \glspl{qubo}, which emphasises the need for efficient classical data structures, representing monomials in \gls{qubo}.
This aspect is of interest in high performance data centres \cite{Wintersperger2022}, where communication time is a vital aspect of performance.
Moreover, estimating time for preparatory steps can be beneficial for schedulers that manage work loads on restricted (quantum) nodes.

The mapping problem (\ie, finding a hardware executable circuit) is also relevant for \gls{qecc}, since they typically add additional gates or qubits to introduce redundancy (see \cite{Roffe_2019, Breuckmann_2021} for more information about \gls{qecc}).
Since circuit depth or the number of additional qubits scales with the number of correctable errors, it is sensible to also consider hardware topology and hardware noise in advance (\eg, by reducing circuit size or depth by a pre-optimised hardware specific mapping).
This aspect can be combined with characteristics that are introduced by transformation paths in an automated quantum toolchain (\eg, connectivity and error rates around additional qubits used for \gls{qecc}).

Overall, our results indicate relatively stable scaling behaviour, paving the way for extrapolation.
Hence, optimising metrics for error correcting codes, hardware and ultimately performance and solution quality is possible in a predictable manner.

\section{Conclusion and Outlook}
\label{sec:ConclAndOutlook}
As there are many transformation paths and free parameters from an abstract problem formulation down to deploying it onto hardware, selecting an optimal route requires the characterisation of global influence of single transformations.
We set the starting point on $k$-\gls{sat} abstraction and investigate four transformation paths down to \gls{qubo}---accompanied by intermediate representations and their respective properties ($3$-\gls{sat}, \gls{mis} and \gls{pubo}).
Our empirical study finds relatively stable property development among all four paths, which enables predicting their behaviour for larger inputs. 
We also give qualitative arguments to extrapolate our findings beyond random $k$-\gls{sat} inputs.
Although properties develop predictably, their magnitude and structure differs (\eg, \# clauses, \# variables, \# monomials, distribution).
An automated quantum toolchain can leverage these differences by optimising for metrics most relevant for a given hardware and therefore improve performance, runtime or solution quality.

The mountainous amount of possible paths and free parameters gives rise to further studies of property development among a wider range of abstraction layers.
For instance, the energy spectra of a resulting optimisation problem (\gls{pubo} or \gls{qubo}) can be interesting in view of (quantum) solvers.
Other studies can also extend our work by incorporating further metrics, ranked by their importance to the target representation.

\newcommand{\LS}{\censor{LS}\xspace}
\newcommand{\WM}{\censor{WM}\xspace}
\newcommand{\ST}{\censor{Simon Thelen}\xspace}
\newcommand{\MF}{\censor{Maja Franz}\xspace}

\newcommand{\programme}{\blackout{German Federal Ministry of
    Education and Research (BMBF), funding program \enquote{Quantum Technologies---from
      Basic Research to Market}}}
\newcommand{\grantoth}{\censor{\#13N15647 and \#13N16092}}
\newcommand{\hta}{\censor{High-Tech Agenda Bavaria}}

\begin{small}
  \noindent\textbf{Acknowledgements}
  We thank \ST for many active discussions and \MF for valuable comments on the ideas presented in this paper. 
  We acknowledge support from \programme, grant \grantoth{}. \WM acknowledges support by the \hta.
\end{small}

\bibliographystyle{IEEEtran}
\bibliography{references.bib}

\end{document}